%% file: MOS_Nf.tex
\documentclass[review]{elsarticle}

\usepackage[utf8]{inputenc}
\usepackage[T1]{fontenc}
\usepackage{tgtermes}
\usepackage{lineno,hyperref,pifont,natbib,geometry,fleqn,graphicx,makeidx,fancyhdr,multirow,verbatim,amsmath,amsthm,amssymb,float,array,subfloat,epstopdf,subfig,upgreek,appendix}
\modulolinenumbers[5]

\journal{Journal of Instrumentation}

\begin{document}

\begin{frontmatter}

\title{
Modeling of Surface Damage at the Si/SiO$_2$-interface of Irradiated MOS-capacitors}


\author[a]{N. Akchurin}
\author[d]{G. Altopp}
\author[d]{B. Burkle}
\author[c]{W. D. Frey}
\author[d]{U. Heintz}
\author[d]{N. Hinton}
\author[s]{M. Hoeferkamp}
\author[a]{Y. Kazhykarim}
\author[b]{V. Kuryatkov}
\author[a]{T. Mengke}
\author[a]{T. Peltola\corref{cor2}}
\ead{timo.peltola@ttu.edu}
\author[s]{S. Seidel}
\author[d]{E. Spencer}
\author[r]{M. Tripathi}
\author[d]{J. Voelker}

\address[a]{Department of Physics and Astronomy, Texas Tech University, 1200 Memorial Circle, Lubbock, U.S.A}
\address[d]{Department of Physics, Brown University, 182 Hope Street, Providence, U.S.A.}
\address[c]{McClellan Nuclear Reactor Center, University of California Davis, 5335 Price Ave, Davis, U.S.A.}
\address[b]{Nanotech Center, Texas Tech University, 902 Boston Ave, Lubbock, U.S.A.}
\address[r]{Physics Department, University of California Davis, 1 Shields Ave, Davis, U.S.A.}
\address[s]{Department of Physics and Astronomy, University of New Mexico, 1919 Lomas Blvd, Albuquerque, U.S.A}



\cortext[cor2]{Corresponding author}




\begin{abstract}
\label{Abstract}
\input{Abstract.tex}

\end{abstract}

%
\end{frontmatter}


\newpage
\section{Introduction}
\label{Introduction}
\input{Introduction.tex}
\section{Samples, effective neutron fluences and doses}
\label{SF}
%
\input{SamplesFluences.tex}

%
\section{Measurement and simulation setups}
\label{Msetup}
%
\input{setups.tex}
\section{Experimental and simulation results}
\label{Results}
%
\input{Results2.tex}
\section{Discussion} 
\label{Discussion}
\input{discussion_fT.tex}

\section{Summary and conclusions}
\label{Conclusions}
%
\input{Summary.tex}
\label{Summary}

\section*{Acknowledgements}
This work has been supported by the US Department of Energy, Office of Science (DE-SC0015592). 
We thank the HGCAL and CMS Tracker collaborations of the CMS Experiment for providing the samples investigated in this study. We thank K. Zinsmeyer, C. Perez, and posthumously P. Cruzan of TTU for their expert technical support. 
%

\bibliography{mybibfile}


\end{document}

%% file: Abstract.tex
Surface damage caused by ionizing radiation in SiO$_2$ passivated silicon particle detectors consists mainly of the accumulation of a positively charged layer 
along with trapped-oxide-charge and interface traps inside the oxide and close to the Si/SiO$_2$-interface. 
High density positive interface net charge can be 
detrimental to the operation of a multi-channel $n$-on-$p$ sensor since the inversion layer generated under the Si/SiO$_2$-interface can cause loss of position 
resolution by creating a conduction channel between the electrodes. 
In the investigation of the radiation-induced accumulation of oxide charge and interface 
traps, a capacitance-voltage characterization study of n/$\gamma$- and $\gamma$-irradiated Metal-Oxide-Semiconductor (MOS) capacitors showed that close agreement between 
measurement and simulation were possible when oxide charge density was complemented by both acceptor- and donor-type deep interface traps with densities comparable 
to the oxide charges. 
Corresponding inter-strip resistance simulations of a $n$-on-$p$ sensor with the tuned oxide charge density and interface traps show close agreement 
with experimental results. 
The beneficial impact of radiation-induced accumulation of deep interface traps on inter-electrode isolation may be considered in the 
optimization of the processing parameters of isolation implants on $n$-on-$p$ sensors for the extreme radiation environments.

%% file: Introduction.tex
In the high-radiation environment of LHC experiments, where silicon particle detectors are utilized,  
radiation-induced defects are introduced both in the silicon substrate (bulk damage or displacement damage) and in
the SiO$_2$ passivation layer, that affect the sensor performance through the interface with the Si-bulk (surface damage). An accompanying study to the investigation presented here 
focused on the influence of displacement damage on the bulk properties of 
test-diode samples exposed to a reactor radiation environment \cite{Peltola2020}. 
In this study, the influence and accumulation of surface damage is investigated by the characterizations of Metal-Oxide-Semiconductor (MOS) capacitors, exposed to either the same reactor
radiation environment as in the previous study or to a $\gamma$-source.

Surface damage caused by ionizing radiation from charged particles, X-rays or gammas in SiO$_2$ passivated devices 
consists of the accumulation of 
positively charged layer (fixed oxide charge density $N_\textrm{f}$ that does not move or exchange charge with Si), 
trapped-oxide-charge and interface traps (or 
surface states 
$N_\textrm{it}$), along with mobile-ionic-charge density ($N_\textrm{M}$) inside the oxide 
and close to the interface with silicon bulk. 
When exposed to high-energy ionizing radiation, electron-hole pairs are created in the oxide. 
The fraction of electrons and holes escaping from the initial recombination drift through the oxide either to the gate-electrode or to the Si/SiO$_2$-interface,
depending on the gate bias. Holes transport with positive gate bias through the oxide toward the Si/SiO$_2$-interface by hopping
through localized states in the oxide, and a fraction of the holes that reach the vicinity of the interface are captured 
by oxygen vacancies (most of the vacancies in the SiO$_2$ are located close to the Si/SiO$_2$-interface) and form positive
oxide-trapped charge. During the transport of holes, some react with hydrogenated oxygen vacancies and result in protons (hydrogen ions). 
The fraction of protons that drift to the interface produce dangling Si-bonds, i.e. interface traps, by breaking the hydrogenated Si-bonds at the interface.
The mechanisms by which surface damage accumulation occurs are 
described extensively in \cite{Nicollian1982,Dressen1989,Oldham1999,Schwank2008,Zhang2012jinst}. 

High positive $N_\textrm{f}$ 
is detrimental to 
the functionality of 
segmented $n$-on-$p$ ($p$-bulk with $n^+$ electrode implants) 
sensors, since the electron layer generated under the Si/SiO$_2$-interface can cause 
loss of position resolution 
by providing a conduction channel that 
compromises the 
inter-electrode isolation. 
Additionally, the accumulating interface charges can contribute to higher channel noise by the increase of inter-electrode capacitance ($C_\textrm{int}$) and modify the electric fields at the edges of the isolation- and channel-implants.
%
$N_\textrm{it}$ can also play a 
notable role in determining the
surface properties of Si-sensors as significant fraction of these 
interface states 
have been reported to be deep traps with densities comparable to $N_\textrm{f}$ \cite{Kim1995,Zhang2013} in X-ray irradiated devices, thus 
capable of altering the space charge near the Si/SiO$_2$-interface. 

Device simulations 
play a vital role in understanding and parametrizing the observed macroscopic effects of the microscopic surface states and charges at the Si/SiO$_2$-interface. 
Previous simulation studies to model the radiation-induced accumulation of surface damage involve several approaches, of which one 
is to approximate the surface damage 
solely by $N_\textrm{f}$ (for the modeling of inter-strip resistance ($R_\textrm{int}$) in Ref. \cite{Verzellesi2000}, $C_\textrm{int}$ in Ref. \cite{Piemonte2006} and electric field surface distribution in Ref. \cite{Unno2011}).
Models complementing $N_\textrm{f}$ with 
interface traps include approaches presented in Table\ref{tabModels}. 
%
%
\begin{table*}[!t]
\centering
\caption{An overview of surface damage models complementing $N_\textrm{f}$ with acceptor- ($N_\textrm{it,acc}$) and/or donor-type ($N_\textrm{it,don}$) $N_\textrm{it}$ at the Si/SiO$_2$-interface.
$E_\textrm{a,V,C}$ are the activation energy, 
valence band and conduction band energies, respectively, while $V_\textrm{fb}$, $R_\textrm{int}$ and $s_\textrm{0}$ are 
the flat-band voltage, inter-strip resistance and surface generation velocity, respectively.} 
\label{tabModels}
\begin{tabular}{|c|c|c|c|}
    \hline
    {\bf $N_\textrm{it}$ type} & {\bf $E_\textrm{a}$} \textrm{[eV]} & {\bf Tuning properties} & {\bf Reference}\\
    \hline
    Deep acceptor & $E_\textrm{C}-0.60$ & \multirow{2}{*}{$V_\textrm{fb}$ and $R_\textrm{int}$} & \multirow{2}{*}{\cite{Dalal2014}}\\
    Shallow acceptor & $E_\textrm{C}-0.39$ & &\\
    \hline
    \hline
    Deep donor & $E_\textrm{V}+0.60$ & \multirow{2}{*}{$V_\textrm{fb}$ and $s_\textrm{0}$} & \multirow{2}{*}{\cite{Morozzi:2020yxm,Morozzi:2021kwo}}\\
    Deep acceptor & $E_\textrm{C}-0.56$ & &\\
    \hline
    \hline
    Deep donor & $E_\textrm{V}+0.70$ & \multirow{3}{*}{ $R_\textrm{int}$} & \multirow{3}{*}{\cite{Moscatelli:2017sps}}\\
    Deep acceptor & $E_\textrm{C}-0.60$ & &\\
    Shallow acceptor & $E_\textrm{C}-0.40$ & &\\
    \hline
\end{tabular}
\end{table*}
A further model, tuned to reproduce the charge-injection position
dependence of charge collection efficiency (CCE($x$)) in the inter-strip region and $C_\textrm{int}$, 
includes an approach where two bulk defect levels tuned for proton irradiation \cite{Eber2013} are augmented by a 
shallow ($E_\textrm{C}-0.40$ eV) $N_\textrm{it,acc}$ with 2 $\upmu$m depth distribution
from the surface \cite{Peltola2015,Peltola2016}. 

It needs to be pointed out that for higher tunability of the models the use of a single set of effective interface traps is an approximation of the real situation, where there can be a continuum of levels at the interface.

The approach taken in this study is to first tune $N_\textrm{f}$ and the parameters of $N_\textrm{it}$ (number of $N_\textrm{it}$, type, energy level, trapping cross-sections and density) to reproduce $CV$-characteristics of irradiated MOS-capacitors and next, as a confidence level test of the tuning, monitor their possible influence on the $R_\textrm{int}$ of a $n$-on-$p$ strip-sensor. The inter-strip isolation results are further examined for the dynamical influence of the fully occupied  $N_\textrm{it}$ on the net $N_\textrm{ox}$ in the operating conditions of reverse biased $n$-on-$p$ sensors.

%
The paper is arranged 
by first introducing the samples, the neutron fluences and the $\gamma$-doses 
the samples 
were exposed to in Section~\ref{SF}. Next, the measurement and simulation setups are descibed in Section~\ref{Msetup}. 
$CV$-characterization results start with the analysis of pre-irradiated reference MOS-capacitors diced from 6-inch wafers in Section~\ref{preIrrad}, where the extracted oxide thicknesses ($t_\textrm{ox}$) and oxide charge densities at the Si/SiO$_2$-interface ($N_\textrm{ox0}$) are presented. 
Results of both reactor neutron and $\gamma$-irradiated 
MOS-capacitors are presented in Section~\ref{Irrad} by first determining the change in the flat band voltage (${\Delta}V_\textrm{fb}$) after irradiation, 
followed by the 
extraction of effective oxide charge density ($N_\textrm{ox}$) at the Si/SiO$_2$-interface. 
Technology Computer-Aided Design (TCAD) simulations are applied in both results sections to reproduce and model the measured $CV$-characteristics. 
The inter-strip resistance 
simulations applying $N_\textrm{f}$ and surface states tuned to reproduce the irradiated MOS-capacitor $CV$-characteristics are presented in the second part of Section~\ref{Irrad}.
Finally, the results are discussed in Section~\ref{Discussion}, while summary and conclusions 
are given in Section~\ref{Summary}. 

%% file: SamplesFluences.tex
Out of five reactor irradiated samples in this study, three were irradiated at
Rhode Island Nuclear Science Center\footnote{http://www.rinsc.ri.gov/} (RINSC) and two at UC Davis McClellan Nuclear Research Center\footnote{https://mnrc.ucdavis.edu/} (MNRC). 
Total Ionizing Dose (TID) in both reactor cores was from a mixed field of neutrons and gammas, while 
the contribution from 
gammas was expected to dominate TID. Monte Carlo N-Particle Transport (MCNP) simulations 
indicated that neutron contribution to TID was only about 5\% at MNRC. In addition, three samples were $\gamma$-irradiated at Sandia National Laboratories Gamma Irradiation Facility\footnote{https://www.sandia.gov/research/gamma-irradiation-facility-and-low-dose-rate-irradiation-facility/} (GIF) with a $^{60}$Co-source. 
After irradiations, the samples were shipped cooled in thermally isolated containers 
to Texas Tech University (TTU) to avoid annealing of the radiation-induced defects and were then 
stored at -$40^{\circ}$ C 
between the measurements.

\begin{figure*}
\centering
\includegraphics[width=.6\textwidth]{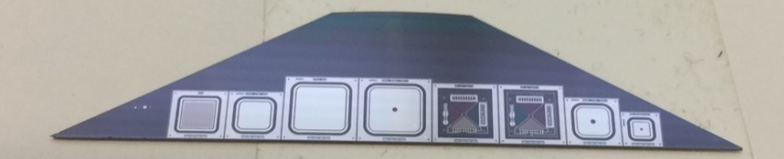}
\caption{\small Half-moon samples with test structures, diced from a 6-inch Si-wafer. The sample has three test diodes, two MOS-capacitors (second and third from left) and other test structures. 
The small and large MOS gate areas are $2.5\times2.5~\textrm{mm}^2$ and $4.0\times4.0~\textrm{mm}^2$, respectively. 
}
\label{Samples}
\end{figure*}

Samples like one shown in Figure~\ref{Samples} were diced from 
6-inch Hamamatsu Photonics K.K. (HPK) 
sensor-wafers 
with $\langle100\rangle$ crystal orientation. The test structures on the samples were designed at Institut f{\"u}r Hochenergiephysik (HEPHY). 
Two distinct diffusion processes were applied in the samples with MOS-capacitors to produce the heavily-doped backplane blocking contact: 
standard-diffusion (`std-FZ') to produce 300-$\upmu\textrm{m}$ active thickness sensors and deep-diffusion (`dd-FZ') for 200- and 120-$\upmu\textrm{m}$ active thicknesses, while the physical thickness of all samples was 320 $\upmu\textrm{m}$. 
Sample identification of e.g. 
`300P$_{-}$std-FZ' in Table~\ref{table_fluence_dose}, 
refers to a 300-$\upmu\textrm{m}$-active thickness MOS-capacitor with $p$-type float-zone Si-bulk and 
standard-diffused backplane implant. Gate oxide in all samples was silicon dioxide (SiO$_2$) with positive oxide charge. 
The 
corresponding effective 1-MeV neutron equivalent fluences ($\Phi_{\textrm{eff}}$) extracted from the test-diode leakage currernts in the samples, $\gamma$-doses and MCNP-simulated TIDs 
are also presented in Table~\ref{table_fluence_dose}.
Details 
of the effective fluence extraction and the bulk properties of the samples are given in \cite{Peltola2020}. 
%
\begin{table*}[!t]
\centering
\caption{Identification of irradiated samples, respective irradiation sites, 
effective neutron fluences ($\Phi_\textrm{eff}$) at RINSC and MNRC, $\gamma$-doses at GIF and MCNP-simulated 
TID at MNRC. 
N1(n): Lowest fluence reactor-irradiated sample with $n$-type bulk before irradiation, G2(p): Second to lowest dose $\gamma$-irradiated sample with $p$-type bulk. Bulk material is indicated as `std-' and `dd-FZ' that correspond to standard- and deep-diffused float zone substrates, respectively. 
}
\label{table_fluence_dose}
\begin{tabular}{|c|c|c|c|c|}
\hline
\multirow{2}{*}{{\bf Sample ID,}} & \multirow{2}{*}{{\bf Irradiation}} & 
\multirow{2}{*}{{\bf $\Phi_\textrm{eff}$}} & \multirow{2}{*}{{\bf D}} & \multirow{2}{*}{{\bf MCNP TID}}\\[0.9mm]
{\bf thickness \& type} & {\bf site} & 
[$\times10^{15}~\textrm{n}_\textrm{eq}$cm$^\textrm{-2}$] & [kGy] & [kGy]\\
\hline
{\bf N1(n):} 300N$_-$std-FZ & RINSC & 
$0.35\pm0.04$ & -- & --\\
\hline 
{\bf N2(n):} 300N$_-$std-FZ & MNRC & 
$0.61\pm0.05$ & -- & $7.1\pm0.6$\\
\hline
{\bf N3(n):} 120N$_-$dd-FZ & RINSC & 
$2.35\pm0.19$ & -- & --\\
\hline
{\bf N4(p):} 120P$_-$dd-FZ & RINSC & 
$6.6\pm0.7$ & -- & --\\
\hline
{\bf N5(n):} 120N$_-$dd-FZ & MNRC & 
$9.3\pm1.1$ & -- & $90\pm11$\\
\hline
{\bf G1(n):} 200N$_-$dd-FZ & GIF & 
-- & $7.0\pm0.4$ &--\\
\hline
{\bf G2(p):} 300P$_-$std-FZ & GIF & 
-- & $23.0\pm1.2$ & --\\
\hline
{\bf G3(n):} 300N$_-$std-FZ & GIF & 
-- & $90\pm5$ & --\\
\hline
\end{tabular}
\end{table*}

%% file: setups.tex
%
The capacitance-voltage ($CV$) characterizations 
were carried out with a dark-box-enclosed custom probe station 
at TTU. 
%
The gate voltage of the MOS-capacitor 
is supplied 
by a Keithley 2410 SMU 
while the 
capacitance is read out by a 
Keysight E4980AL LCR-meter. The 
decoupling of the LCR-meter's high- and low-terminals from the DC-circuit 
is 
accomplished by 1-$\upmu\textrm{F}$ capacitors. The remote control and 
data acquisition functions are carried out with LabVIEW\texttrademark-based software.
%
The MOS-capacitor 
is connected to the measurement circuit by a vacuum chuck from its backplane, which also provides fixed position, and by a probe needle on the segmented front surface. 

Simulations 
were carried out using the Synopsys Sentaurus\footnote{http://www.synopsys.com} finite-element 
TCAD software framework. 
The parameters extracted from $CV$-measurements of the MOS-capacitors (oxide thickness ($t_\textrm{ox}$), 
flat band capacitance ($C_\textrm{fb}$) and voltage ($V_\textrm{fb}$)) and test diodes on the same samples (bulk doping ($N_\textrm{B}$) and backplane deep-diffusion doping profiles (see ref. \cite{Peltola2020} for details)) before irradiation, as well as their known gate area, were used as inputs to reproduce the devices as closely as possible by a 2D-simulation. 

%% file: Results2.tex
\subsection{Results before irradiation}
\label{preIrrad}
\begin{figure*}
     \centering
     \subfloat[ ]{\includegraphics[width=.495\textwidth]{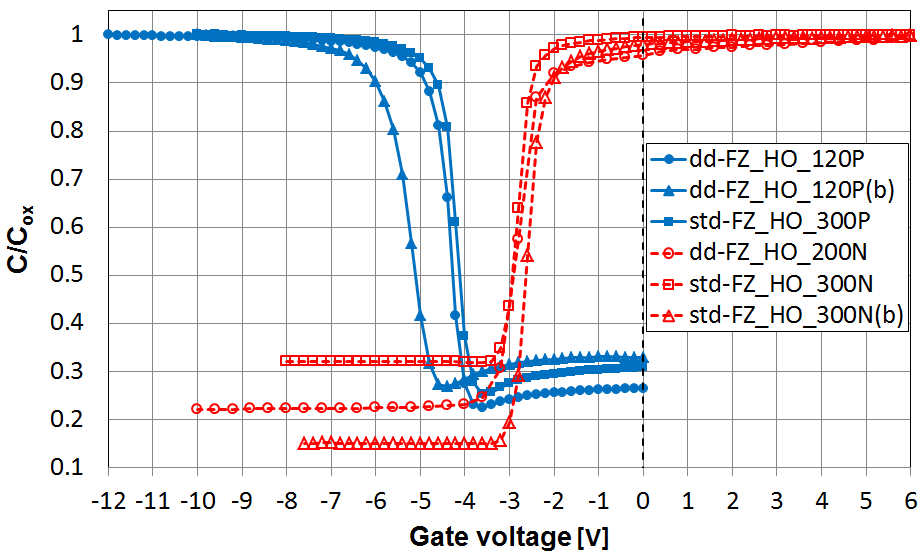}\label{CV_npMOS}}\hspace{1mm}%
     \subfloat[ ]{\includegraphics[width=.495\textwidth]{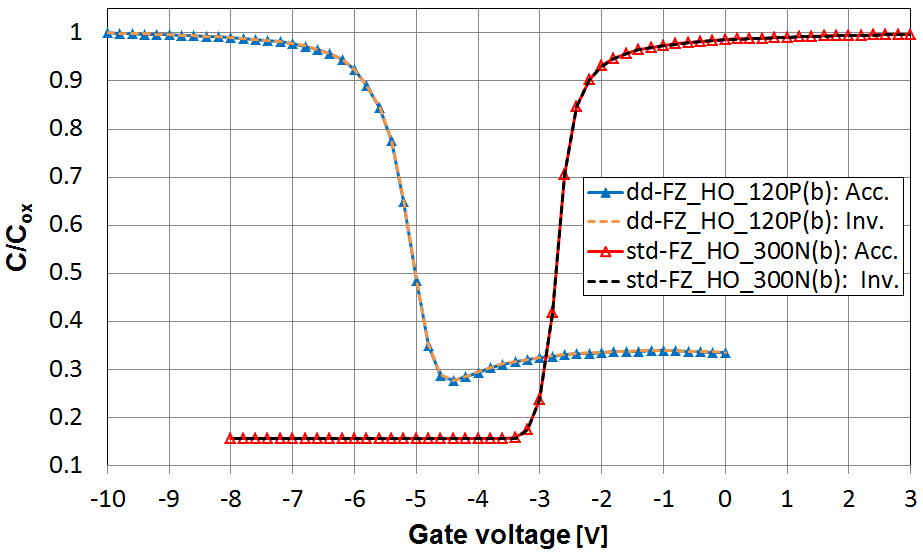}\label{CV_npMOS_hyst}}\\
     \subfloat[ ]{\includegraphics[width=.55\textwidth]{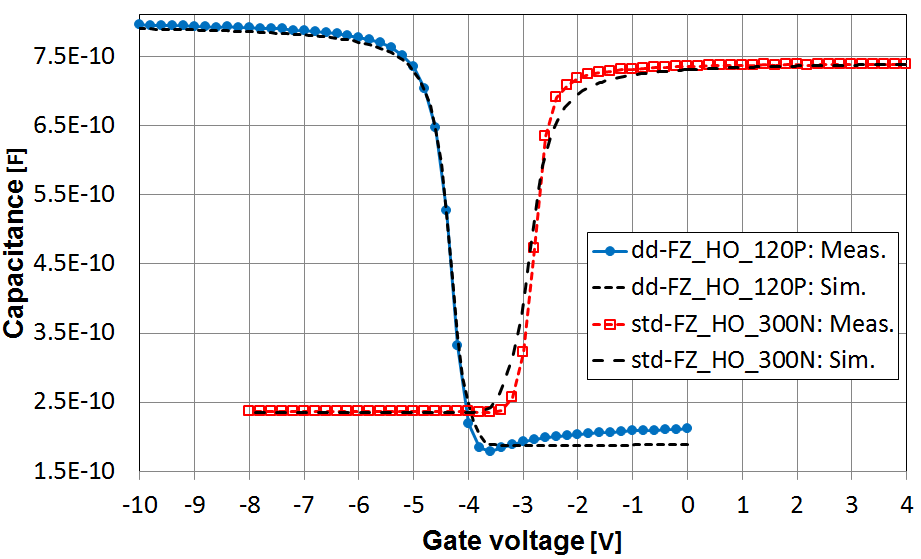}\label{CVsim6001_1102}}
    \caption{(a) Measured $C_\textrm{ox}$-normalized $CV$-characteristics of six pre-irradiated reference MOS-capacitors with both $p$- and $n$-type bulk materials. 
    (b). Measured $C_\textrm{ox}$-normalized $CV$-characteristics of $p$- and $n$-bulk reference MOS-capacitors with $CV$-sweep starting either from accumulation (Acc.) or inversion (Inv.) regions. (c) Measured and simulated $CV$-characteristics of $p$- and $n$-bulk reference MOS-capacitors at $T=293$ K and $f=200$ kHz. 
}
\label{CV_npMOS_simMeas}
\end{figure*}

The 
$CV$-measurements of six 
reference MOS-capacitors were carried out at room temperature and the resulting $CV$-characteristics are presented in Figure~\ref{CV_npMOS}, where the three 
modes of operation --accumulation, depletion and inversion-- 
are visible. 
At gate voltages with $C/C_\textrm{ox}=1$, where $C_\textrm{ox}$ is the oxide capacitance, 
the majority carriers (holes for $p$-bulk, electrons for $n$-bulk) are pulled to the Si/SiO$_2$-interface, forming an accumulation layer with zero surface potential. An abrupt drop of 
capacitance takes place in the depletion region, where the Si-surface is being depleted from majority carriers and measured $C$ is now $C_\textrm{ox}$ and depletion layer capacitance in series. Thus, the measured $C$ keeps decreasing with gate voltage as the effective thickness of the depletion region, that acts as a dielectric between the gate and the Si-substrate, increases. Depth of the depletion 
region reaches its maximum, and measured $C$ its minimum ($C_\textrm{min}$), when most of the available minority carriers are pulled to the Si/SiO$_2$-interface (by the positive oxide charge 
and by the negative gate voltage for $p$- and $n$-bulk MOS-capacitors, respectively, in Figure~\ref{CV_npMOS}), forming an inversion layer. The depletion region in the measured $CV$-curve is limited by the threshold voltage ($V_\textrm{th}$), where the surface potential equals twice the bulk potential, and the flat band voltage ($V_\textrm{fb}$), where the Si energy band becomes flat and the surface potential goes to zero \cite{Nicollian1982}.

Oxide thicknesses were extracted by the relation
\begin{linenomath}
\begin{equation}\label{eq1}
t_\textrm{ox}=\epsilon_\textrm{ox}\frac{A}{C_\textrm{ox}},
\end{equation}
%
where 
$\epsilon_\textrm{ox}$ is the product of vacuum permittivity and the relative permittivity of the 
oxide material (SiO$_2$), and 
$A$ is the MOS gate area 
\cite{Sze1981}. Flat band voltages were determined by the flat-band-capacitance method with
\begin{equation}\label{eq2}
C_\textrm{fb}=\frac{C_\textrm{ox}\epsilon_\textrm{S}A/\lambda}{C_\textrm{ox}+\epsilon_\textrm{S}A/\lambda},
\end{equation}
where $\epsilon_\textrm{S}$ is the permittivity of the substrate material (Si) and $\lambda$ is the extrinsic Debye length \cite{Nicollian1982}. Flat band voltage $V_\textrm{fb}$ is extracted from the $CV$-curve at the value of $C_\textrm{fb}$. 
Effective oxide charge densities at the Si/SiO$_2$-interface were then calculated by
\begin{equation}\label{eq3}
N_\textrm{ox}=\frac{C_\textrm{ox}}{eA}(W_\textrm{MS}-V_\textrm{fb}),
\end{equation}
where $W_\textrm{MS}$ is the metal-semiconductor work function 
and $e$ is the elementary charge. The results for $t_\textrm{ox}$, $V_\textrm{fb0}$ and $N_\textrm{ox0}$ of the reference MOS-capacitors are presented in Table~\ref{table_preIrrad}. $N_\textrm{ox}$ extracted from $CV$-measurements can be stated as
\begin{equation}\label{eq3b}
N_\textrm{ox}=N_\textrm{f}+N_\textrm{it}+N_\textrm{M},
\end{equation}
\end{linenomath}
where the total value of $N_\textrm{ox}$ 
is influenced by 
fixed-oxide-charge ($N_\textrm{f}$), interface-trapped-charge ($N_\textrm{it}$) and mobile-ionic-charge ($N_\textrm{M}$) densities 
\cite{Nicollian1982}. $N_\textrm{M}$ can be observed in $CV$-curve hysteresis when the charges (typically alkali metal ions such as Na$^+$, K$^+$ and Li$^+$ before irradiation \cite{Greeuw1984}) drift between metal/SiO$_2$- and Si/SiO$_2$-interfaces under the applied electric field, depending on the gate bias and the polarity of the charges \cite{Nicollian1982,Suria2015}. The results in Figure~\ref{CV_npMOS_hyst} do not show any sign of hysteresis. Scanned frequencies involved 1$-$200 kHz, 
while the low mobility of $N_\textrm{M}$ in oxide (highest for Na$^+$ with $\mu\approx4\times10^{-12}$ cm$^2$/Vs at room temperature \cite{Nicollian1982}) 
that results in long drift time from metal- to Si-interface (about 2 min with $E\approx10^5$ V/cm and $t_\textrm{ox}\approx700$ nm), was taken into account by applying extended gate-voltage hold-time at strong inversion and strong accumulation before $CV$-sweep, respectively. Thus, the influence of $N_\textrm{M}$ in the reference samples is considered negligible, reducing Eq.~\ref{eq3b} to $N_\textrm{ox}{\cong}N_\textrm{f}+N_\textrm{it}$. 
This is in contrast with previous results for thin ($<$25 nm) 
SiO$_2$ films, where significant influence of $N_\textrm{M}$ was observed \cite{Suria2015}. 
The oxide films in Table~\ref{table_preIrrad} 
$-$produced by high-growth-rate deposition method 
like wet thermal oxidation \cite{Oviroh2019,Jaeger2002}$-$ are about 30 times thicker than the oxide films in ref. \cite{Suria2015}.
Increased trapping probability of ions with substantially longer drift distance from metal- 
to Si-interface 
could explain the lack of influence from $N_\textrm{M}$ in the results.

The measured $CV$-characteristics in Figure~\ref{CVsim6001_1102} were closely reproduced by TCAD simulation. 
Matching $V_\textrm{fb}$-values within uncertainty to the measured were reached in the simulation by using $N_\textrm{f}$ (implemented as a positive charge-sheet located at the Si/SiO$_2$-interface with a uniform distribution along the interface) as the sole tuning parameter. Frequency-scan $f=1-200$ kHz showed no frequency dependence on the measured value of $V_\textrm{fb}$, which was repeated by the simulation. The values of $N_\textrm{ox}$ (extracted from the measurement) and $N_\textrm{f}$ (required in the simulation) for 120P MOS-capacitor in Figure~\ref{CVsim6001_1102} were $(1.120\pm0.012)\times10^{11}~\textrm{cm}^{-2}$ and $(1.14\pm0.02)\times10^{11}~\textrm{cm}^{-2}$, respectively, while the corresponding values for 300N were $(6.67\pm0.02)\times10^{10}~\textrm{cm}^{-2}$ and $(7.00\pm0.03)\times10^{10}~\textrm{cm}^{-2}$, respectively. In four measured and simulated reference MOS-capacitors the required $N_\textrm{f}$ 
deviated from the measured $N_\textrm{ox}$ by $+2.8\pm1.5\%$. 

Thus, the combined hysteresis and simulation results show evidence that $N_\textrm{ox}$ in the reference samples is essentially governed by $N_\textrm{f}$, i.e. $N_\textrm{ox}{\cong}N_\textrm{f}$. This is in line 
with the results of ref. \cite{Poehlsen2013}, where it was reported that 
$N_\textrm{it}$ at the 
Si/SiO$_2$-interface of a non-irradiated silicon device is about two-orders-of-magnitude lower than 
$N_\textrm{f}$ at the interface. 
Therefore, 
the density of surface traps corresponding to $N_\textrm{f}=1\times10^{11}~\textrm{cm}^{-2}$ 
should be in the order of $10^{9}~\textrm{cm}^{-2}$. Additional simulation with 
surface-trapping-center energy of $E_\textrm{a}=E_\textrm{c}-0.60~\textrm{eV}$ \cite{Zhang2012,Peltola2015r} and a fairly large capture cross-section of 
$1\times10^{-14}~\textrm{cm}^{2}$ results in identical $CV$-characteristics to the simulations in Figure~\ref{CVsim6001_1102} until the difference between $N_\textrm{it}$ and $N_\textrm{f}$ is less than one-order-of-magnitude.

%
\begin{table*}[!t]
\centering
\caption{Reference sample identification, measured oxide thicknesses ($t_\textrm{ox}$), flat band voltages ($V_\textrm{fb0}$) and the 
effective oxide charge densities ($N_\textrm{ox0}$) at the Si/SiO$_2$-interface 
before irradiation. 
}
\label{table_preIrrad}
\begin{tabular}{|c|c|c|c|}
\hline
\multirow{2}{*}{{\bf Sample thickness \&}} & 
\multirow{2}{*}{{\bf $t_\textrm{ox}$}} & \multirow{2}{*}{{\bf $V_\textrm{fb0}$}} & \multirow{2}{*}{{\bf $N_\textrm{ox0}$}}\\[0.5mm]
{\bf type} & 
[nm] & [-V] & [$\times10^{10}~\textrm{cm}^{-2}$]\\
\hline
120P$_-$dd-FZ & 
$695\pm5$ & $4.38\pm0.03$ & $11.20\pm0.12$\\
\hline
120P$_-$dd-FZ(b) & 
$687\pm5$ & $5.35\pm0.04$ & $14.30\pm0.14$\\
\hline
300P$_-$std-FZ & 
$795\pm4$ & $4.24\pm0.02$ & $9.39\pm0.06$\\
\hline
200N$_-$dd-FZ & 
$741\pm16$ & $2.69\pm0.06$ & $5.59\pm0.17$\\
\hline
300N$_-$std-FZ & 
$749\pm2$ & $2.750\pm0.007$ & $6.67\pm0.02$\\
\hline
300N$_-$std-FZ(b) & 
$704\pm4$ & $2.600\pm0.015$ & $6.54\pm0.05$\\
\hline
\end{tabular}
\end{table*}
%
\subsection{Results after n/$\gamma$- and $\gamma$-irradiations}
\label{Irrad}
\begin{figure*}
     \centering
     \subfloat[]{\includegraphics[width=.56\textwidth]{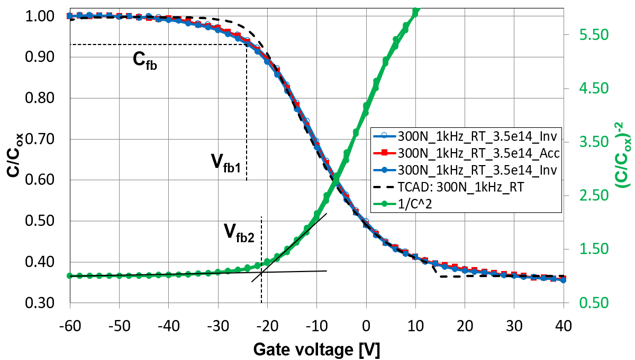}\label{300N_1kHz_35e14}}\\
     \subfloat[]{\includegraphics[width=.5\textwidth]{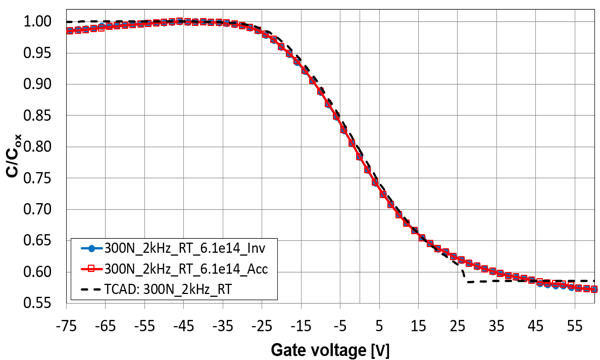}\label{300N_2kHz_61e14}}
     \subfloat[]{\includegraphics[width=.5\textwidth]{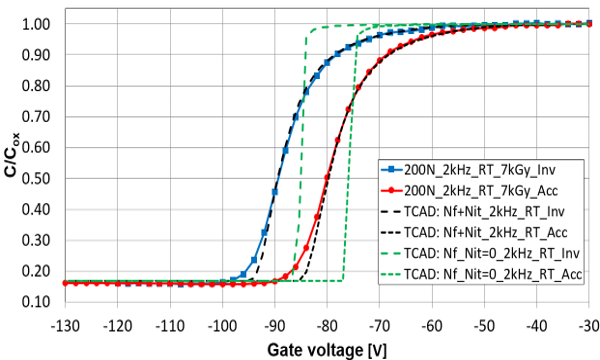}\label{200N_2kHz_7kGy}}
    \caption{
    Measured and simulated $CV$-characteristics at $T=\textrm{RT}$ 
    for a (a) 300-$\upmu$m-thick initial $n$-bulk (type inverted to $p$-bulk) MOS-capacitor (ID: N1(n)) n/$\gamma$-irradiated to the fluence of $(3.5\pm0.4)\times10^{14}~\textrm{n}_\textrm{eq}\textrm{cm}^{-2}$ and measured at $f=1$ kHz (with the two methods to extract measured $V_\textrm{fb}$ displayed), (b) 300-$\upmu$m-thick initial $n$-bulk (type inverted to $p$-bulk) MOS-capacitor (ID: N2(n)) n/$\gamma$-irradiated to the fluence of $(6.1\pm0.5)\times10^{14}~\textrm{n}_\textrm{eq}\textrm{cm}^{-2}$ (and to the MCNP-simulated $\textrm{TID}=7.1\pm0.6$ kGy) and measured at $f=2$ kHz, and (c) 200-$\upmu$m-thick $n$-bulk MOS-capacitor (ID: G1(n)) $\gamma$-irradiated to the dose of $7.0\pm0.4$ kGy and measured at $f=2$ kHz. 
    Measurements included voltage-scans starting from both inversion (Inv.) and accumulation (Acc.) regions. Simulation input parameters in Figures~\ref{f_1_2kHz}--\ref{f_3kHz} are presented in Tables~\ref{tabNit} and~\ref{table_Simulated}. 
   } 
\label{f_1_2kHz}
\end{figure*}

\begin{figure*}
     \centering
     \subfloat[]{\includegraphics[width=.5\textwidth]{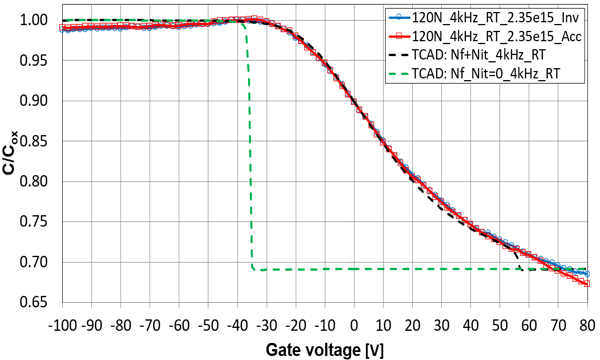}\label{120N_4kHz_235e15}}
     \subfloat[]{\includegraphics[width=.5\textwidth]{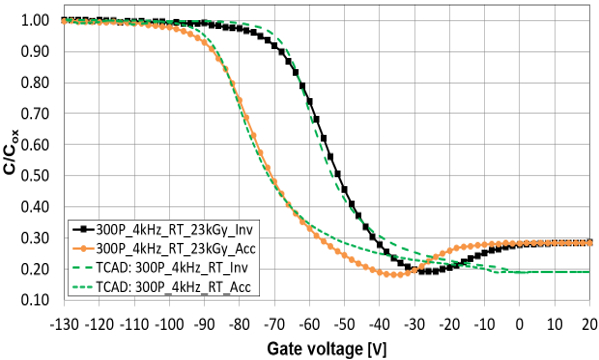}\label{300P_4kHz_23kGy}}
    \caption{Measured and simulated $CV$-characteristics at $T=\textrm{RT}$ and $f=4$ kHz for a (a) 120-$\upmu$m-thick initial $n$-bulk (type inverted to $p$-bulk) MOS-capacitor (ID: N3(n)) n/$\gamma$-irradiated to the fluence of $(2.35\pm0.19)\times10^{15}~\textrm{n}_\textrm{eq}\textrm{cm}^{-2}$, and (b) 300-$\upmu$m-thick $p$-bulk MOS-capacitor (ID: G2(p)) $\gamma$-irradiated to the dose of $23.0\pm1.2$ kGy. Measurements included voltage-scans starting from both inversion (Inv.) and accumulation (Acc.) regions. 
   } 
\label{f_4kHz}
\end{figure*}

\begin{figure*}
     \centering
     \includegraphics[width=.5\textwidth]{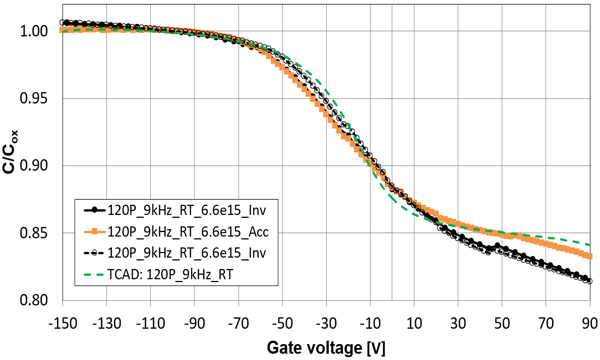}\label{120P_9kHz_66e15}
    \caption{Measured and simulated $CV$-characteristics at $T=\textrm{RT}$ and $f=9$ kHz for 120-$\upmu$m-thick $p$-bulk MOS-capacitor n/$\gamma$-irradiated to the fluence of $(6.6\pm0.7)\times10^{15}~\textrm{n}_\textrm{eq}\textrm{cm}^{-2}$ (ID: N4(p)). 
    Measurements included voltage-scans starting from both inversion (Inv.) and accumulation (Acc.) regions.  
   } 
\label{f_9kHz}
\end{figure*}

The $CV$-measurements of five n/$\gamma$-irradiated and three $\gamma$-irradiated 
MOS-capacitors were carried out at room temperature with frequencies 1--9 kHz, 
and the resulting $CV$-characteristics are presented in 
Figures~\ref{f_1_2kHz}--\ref{f_9kHz}. Due to the extended depletion-regions and the gradual change from depletion to accumulation in the $CV$-curves, most probable estimates for the $V_\textrm{fb}$ of the irradiated samples 
were extracted by both $C_\textrm{fb}$-method in Eq.~\ref{eq2} and from the crossing-point of linear fits to the dynamic (depletion) and static (accumulation) regions of the reciprocal $C^2$-curve, as shown in Figure~\ref{300N_1kHz_35e14}.
The final value and the uncertainty of $V_\textrm{fb}$ 
was then given by the average and the difference of the two extracted values, respectively. The results for $V_\textrm{fb}$ and its relative change 
before and after irradiation (${\Delta}V_\textrm{fb}$) are presented in Table~\ref{table_Irrad}.  

\begin{linenomath}
The change of $N_\textrm{ox}$ with ionizing radiation 
is proportional to the change of $W_\textrm{MS}$ and $V_\textrm{fb}$ in Eq.~\ref{eq3} before ($W_\textrm{MS0}$ and $V_\textrm{fb0}$) and after 
irradiation
\begin{equation}\label{eq4a}
N_\textrm{ox}-N_\textrm{ox0}=\frac{C_\textrm{ox}}{eA}(W_\textrm{MS}-W_\textrm{MS0}-(V_\textrm{fb}-V_\textrm{fb0})).
\end{equation}
%
The results of the bulk properties of the same reactor-irradiated samples in the accompanying study \cite{Peltola2020} indicate that the effective bulk doping ($N_\textrm{eff}$) varies in the studied fluence range from a few times $10^{12}~\textrm{cm}^{-3}$ (before irradiation) to close to $1\times10^{14}~\textrm{cm}^{-3}$ (at the highest fluence of about $1\times10^{16}~\textrm{n}_\textrm{eq}\textrm{cm}^{-2}$). The $N_\textrm{eff}$ 
dependence of $W_\textrm{MS}$ leads in this range to the maximal change of ${\Delta}W_\textrm{MS,max}\approx0.12~\textrm{V}$, regardless of bulk polarity \cite{Nicollian1982}. In comparison, the values of ${\Delta}V_\textrm{fb}$ in Table~\ref{table_Irrad} are two-orders-of-magnitude higher to ${\Delta}W_\textrm{MS,max}$. Thus, contribution from ${\Delta}W_\textrm{MS}$ to $N_\textrm{ox}$ is considered negligible in the following analysis, simplifying Eq.~\ref{eq4a} to
\begin{equation}\label{eq4}
N_\textrm{ox}\cong\frac{C_\textrm{ox}}{eA}(V_\textrm{fb0}-V_\textrm{fb})+N_\textrm{ox0}=\frac{C_\textrm{ox}}{eA}{\Delta}V_\textrm{fb}+N_\textrm{ox0}.
\end{equation}
Since the absolute values of $C_\textrm{ox}$ in the n/$\gamma$-irradiated MOS-capacitors were observed to fluctuate significantly from the initial $C_\textrm{ox}$ before irradiation, the relation applied to extract $N_\textrm{ox}$ was reached by combining Eqs.~\ref{eq1} and~\ref{eq4}:
\begin{equation}\label{eq5}
N_\textrm{ox}\cong\frac{\epsilon_\textrm{ox}}{t_\textrm{ox}e}{\Delta}V_\textrm{fb}+N_\textrm{ox0},
\end{equation}
%
where $N_\textrm{ox}$ is solely dependent on oxide thickness, change in the flat band voltage and the initial oxide charge density at Si/SiO$_2$-interface before irradiation. The results for $N_\textrm{ox}$ 
are included in Table~\ref{table_Irrad}.
\end{linenomath}
%
\begin{table*}[!t]
\centering
\caption{Identification of the irradiated samples, 
total ionizing doses (TID), measured flat band voltages ($V_\textrm{fb}$), relative change in 
$V_\textrm{fb}$ before and after irradiation (${\Delta}V_\textrm{fb}$) and 
Si/SiO$_2$-interface oxide charge densities ($N_\textrm{ox}$) after irradiation. 
Corresponding $CV$-sweep starting regions for inversion (Inv.) and accumulation (Acc.) are also indicated. 
The TID-values in parentheses are interpolations from MCNP-simulated TIDs. 
}
\label{table_Irrad}
\begin{tabular}{|c|c|c|c|c|c|}
\hline
\multirow{2}{*}{{\bf Sample}} & 
\multirow{2}{*}{{\bf TID}} & \multirow{2}{*}{{\bf $CV$-sweep}} & \multirow{2}{*}{{\bf $V_\textrm{fb}$}} & \multirow{2}{*}{{\bf ${\Delta}V_\textrm{fb}$}} & \multirow{2}{*}{{\bf $N_\textrm{ox}$}}\\[0.9mm]
{\bf ID} & 
[kGy] & { } & [-V] & [V] & [$\times10^{12}~\textrm{cm}^{-2}$]\\
\hline
{\bf N1(n)} & 
-- & Inv./Acc. & $22.7\pm1.5$ & $20.0\pm1.3$ & $0.64\pm0.04$\\
\hline 
{\bf N2(n)} & 
$7.1\pm0.6$ & Inv./Acc. & $26\pm3$ & $23\pm3$ & $0.77\pm0.10$\\
\hline
{\bf N3(n)} & 
($23.5\pm1.9$) & Inv./Acc. & $37\pm4$ & $34\pm3$ & $1.06\pm0.15$\\
\hline
{\bf N4(p)} & 
($64\pm7$) & Inv./Acc. & $65\pm4$ & $59\pm4$ & $2.00\pm0.14$\\
\hline
{\bf N5(n)} & 
$90\pm11$ & Inv./Acc. & $71\pm6$ & $68\pm6$ & $2.1\pm0.3$\\
\hline
\multirow{2}{*}{{\bf G1(n)}} & 
\multirow{2}{*}{$7.0\pm0.4$} & Inv. & $86\pm4$ & $83\pm4$ & $2.47\pm0.15$\\
& & Acc. & $76\pm4$ & $73\pm4$ & $2.19\pm0.14$\\
\hline
\multirow{2}{*}{{\bf G2(p)}} & 
\multirow{2}{*}{$23.0\pm1.2$} & Inv. & $62\pm7$ & $58\pm7$ & $1.66\pm0.19$\\
& & Acc. & $81\pm8$ & $77\pm7$ & $2.17\pm0.20$\\
\hline
\multirow{2}{*}{{\bf G3(n)}} & 
\multirow{2}{*}{$90\pm5$} & Inv. & $111\pm14$ & $108\pm14$ & $3.2\pm0.4$\\
& & Acc. & $85\pm13$ & $82\pm12$ & $2.4\pm0.4$\\
\hline
\end{tabular}
\end{table*}

\begin{figure*}
     \centering
     \subfloat[]{\includegraphics[width=.5\textwidth]{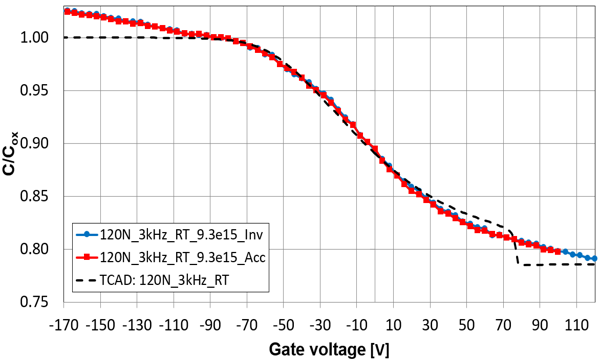}\label{120N_3kHz_93e15}}
     \subfloat[]{\includegraphics[width=.5\textwidth]{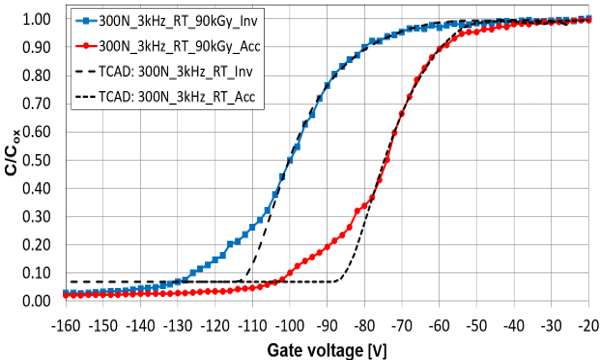}\label{300N_3kHz_90kGy}}
    \caption{Measured and simulated $CV$-characteristics at $T=\textrm{RT}$ and $f=3$ kHz for a (a) 120-$\upmu$m-thick initial $n$-bulk (type inverted to $p$-bulk) MOS-capacitor (ID: N5(n)) n/$\gamma$-irradiated to the fluence of $(9.3\pm1.1)\times10^{15}~\textrm{n}_\textrm{eq}\textrm{cm}^{-2}$ (and to the MCNP-simulated $\textrm{TID}=90\pm11$ kGy), and (b) 300-$\upmu$m-thick $n$-bulk MOS-capacitor (ID: G3(n)) $\gamma$-irradiated to the dose of $90\pm5$ kGy. Measurements included voltage-scans starting from both inversion (Inv.) and accumulation (Acc.) regions. 
   } 
\label{f_3kHz}
\end{figure*}

As displayed by the measurements in Figures~\ref{300N_1kHz_35e14},~\ref{300N_2kHz_61e14},~\ref{120N_4kHz_235e15},~\ref{f_9kHz} and~\ref{120N_3kHz_93e15}, $CV$-characteristics of the n/$\gamma$-irradiated MOS-capacitors are distinctly different compared to the results observed before irradiation. The about 1 V-wide depletion regions displayed in Figure~\ref{CV_npMOS_simMeas} have increased to tens of V 
and $V_\textrm{fb}$ has an opposite sign compared to 
$V_\textrm{th}$, while hysteresis between $CV$-sweeps starting either from accumulation or inversion regions is absent. 
Additionally, the inversion region has a non-zero slope that 
is independent of measurement and delay times, and of $CV$-sweep direction, while 
the accumulation region in the initally $n$-bulk MOSs 
has been reversed to opposite sign gate-voltages ($V_\textrm{gate}$) from Figure~\ref{CV_npMOS_simMeas}, indicating neutron-irradiation-induced space charge sign inversion (SCSI), where $n$-type Si-bulk is turned effectively to $p$-type by the introduction of acceptor-type bulk defects, 
switching the majority carriers 
from electrons to holes. Further signs of the neutron-irradiation-induced displacement damage in the Si-substrate 
is displayed by the significant increase in the ratio $C_\textrm{min}/C_\textrm{ox}$ \cite{Fernandez2013,Luna2006} (from $\leq$0.3 before irradiation to $\leq$0.83 after irradiation, increasing with fluence). 
The accumulation of displacement damage with neutron irradiation in the Si-bulk of the samples is considered only qualitatively in the scope of this study, while more detailed investigation 
is presented in \cite{Peltola2020}. 
Finally, to produce stable $CV$-curves from the n/$\gamma$-irradiated MOSs, careful tuning of the measurement frequency was required, resulting in specific $f$ (1--9 kHz) for each of the five investigated fluences. The three $\gamma$-irradiated MOS-capacitors in the study were exposed to doses within uncertainty to the corresponding TID-values of the three neutron irradiated MOSs in Table~\ref{table_Irrad}, and were measured at identical frequencies to the n/$\gamma$-irradiated MOSs to enable meaningful comparison.

Although the threshold energy for X-rays or gammas to induce bulk damage has been reported to be about 300 keV \cite{Zhang2012jinst} ($E_{\gamma}(^{60}\textrm{Co})=1.173, 1.332$ MeV), none of the aforementioned bulk damage effects are visible on the $CV$-characteristics of the MOS-capacitors irradiated solely by gammas in Figures~\ref{200N_2kHz_7kGy},~\ref{300P_4kHz_23kGy} 
and~\ref{300N_3kHz_90kGy}. The depletion region has again expanded to tens of V, as for n/$\gamma$-irradiated MOSs, but $V_\textrm{th}$ and $V_\textrm{fb}$ are now of the same sign, while the $CV$-curves starting from the accumulation or inversion regions display significant hysteresis ($\geq$10 V) that increases with dose. 
In Figures~\ref{200N_2kHz_7kGy},~\ref{300P_4kHz_23kGy} and~\ref{300N_3kHz_90kGy} higher absolute value of $V_\textrm{fb}$ (= higher positive $N_\textrm{ox}$) is always seen when $CV$-sweep starts from negative $V_\textrm{gate}$. This rules out the influence of $N_\textrm{M}$, that at negative $V_\textrm{gate}$ would always reduce the net value of positive $N_\textrm{ox}$, regardless of the charge sign of $N_\textrm{M}$. 

The comparison between the $\gamma$-irradiated MOS-capacitors with doses within uncertainty to the corresponding MCNP-simulated/interpolated TID-values of the three n/$\gamma$-irradiated MOSs in Table~\ref{table_Irrad} shows systematically lower $V_\textrm{fb}$ and $N_\textrm{ox}$-values for the n/$\gamma$-irradiated MOSs, suggesting that the simulated TID-levels at MNRC reactor were overestimated. 

To investigate and reproduce the 
$CV$-characteristics of the irradiated MOS-capacitors in Figures~\ref{f_1_2kHz}--\ref{f_3kHz} by TCAD simulation, MOS-structures with 
input parameters described in Section~\ref{Msetup} were applied. For the n/$\gamma$-irradiated MOSs, the occurance of SCSI in $n$-type bulk and the accumulation of the effective bulk doping with neutron fluence were approximated by using $p$-type bulk and 
tuning $N_\textrm{eff}$ 
to reach similar $C_\textrm{min}$/$C_\textrm{ox}$-ratios with the measurements, while for $\gamma$-irradiated MOSs, in the absence of displacement damage in the bulk, both $n$- and $p$-bulks were utilized. Simulations 
were carried out at matching temperature and frequencies to the measurements. 

\begin{linenomath}
As shown in Figures~\ref{200N_2kHz_7kGy} and~\ref{120N_4kHz_235e15}, including only $N_\textrm{f}$ (within uncertainty to the measured $N_\textrm{ox}$ in Table~\ref{table_Simulated}) 
at the Si/SiO$_2$-interface 
results in an abrupt depletion region that does not reproduce the characteristics of the measured $CV$-curves of either n/$\gamma$- or $\gamma$-irradiated MOS-capacitors. 
Hence, close agreement 
between measured and simulated $CV$-characteristics was found to require the introduction of both donor- and acceptor-type $N_\textrm{it}$ ($N_\textrm{it,don}$ and $N_\textrm{it,acc}$, respectively) at the 
interface. Data from previous studies on X-ray irradiated MOS-capacitors and strip sensors \cite{Zhang2012jinst,Zhang2013,Poehlsen2013,Zhang2012,Zhang2012synch,Poehlsen2013b,PohlsenPHD2013} was used as a starting point for setting the parameter values of $N_\textrm{it}$ in Table~\ref{tabNit}, and for further tuning, the surface-state dynamics \cite{Nicollian1982}, 
\begin{equation}\label{eq6}
\begin{aligned}
&N_\textrm{f}: \textrm{always fully occupied,}\\%
&\textrm{unoccupied}~N_\textrm{it,don}~\&~N_\textrm{it,acc}: Q_\textrm{it}=0,\\%
&\textrm{fully occupied}~N_\textrm{it,don}: Q_\textrm{it}=+e,~\textrm{and}\\%
&\textrm{fully occupied}~N_\textrm{it,acc}: Q_\textrm{it}=-e,
\end{aligned}
\end{equation}
where $Q_\textrm{it}$ is the interface-trapped charge and $e$ is the elementary charge, were considered.
\end{linenomath}

The fully occupied 
$N_\textrm{it,acc}$ 
increase the electron density at the 
Si/SiO$_2$-interface, 
reducing the net positive oxide charge at the interface, eventually moving $V_\textrm{th}$ from negative to positive 
$V_\textrm{gate}$ with the tuned parameter values presented in Tables~\ref{table_Simulated} and~\ref{tabNit}. The fully occupied 
$N_\textrm{it,don}$ keep the 
hole density at the interface 
high enough to widen the depletion region over an 
extended voltage range 
with a gradual 
slope. 
After the tuning of the parameter values for agreeing 
$CV$-characteristics with the measurements, 
the densities of both types of $N_\textrm{it}$ in Table~\ref{table_Simulated} remain comparable to the magnitude of $N_\textrm{f}$, which is in line with previous experimental observations \cite{Kim1995,Zhang2013,Dalal2014}. Additionally, to reproduce the measured $CV$-characteristics, it was found to be necessary that both $N_\textrm{it}$ are deep traps, 
since tuning of one or both $N_\textrm{it}$ with shallow (0.39 eV and 0.48 eV) traps reported in \cite{Zhang2012synch} did not produce agreement with the measurements. Also, the $E_\textrm{a}$ of $N_\textrm{it,don}$ needed to be tuned from the initial $E_\textrm{V}+0.60$ eV to $E_\textrm{V}+0.65$ eV in Table~\ref{tabNit} for a closer match with the experimental results. The trapping cross-section $1\times10^{-15}$ cm$^2$ for deep traps in \cite{Zhang2012synch} was not tuned from the initial value for both electrons and holes.
\begin{table*}[!t]
\centering
\caption{Irradiated sample identification, $CV$-sweep starting regions for inversion (Inv.) and accumulation (Acc.), measured Si/SiO$_2$-interface oxide charge densities ($N_\textrm{ox}$) after irradiation, and TCAD-simulation input densities for fixed oxide charge ($N_\textrm{f}$) and donor- and acceptor-type interface traps ($N_\textrm{it,don}$ \& $N_\textrm{it,acc}$, respectively) from Table~\ref{tabNit} to reproduce measured $CV$-characteristics in Figures~\ref{f_1_2kHz}--\ref{f_3kHz}. 
}
\label{table_Simulated}
\begin{tabular}{|c|c|c|c|c|c|}
\hline 
\multirow{2}{*}{{\bf Sample}} & \multirow{2}{*}{{\bf $CV$-sweep}} & \multirow{2}{*}{{\bf $N_\textrm{ox}$}} & \multirow{2}{*}{{\bf TCAD $N_\textrm{f}$}} & \multirow{2}{*}{{\bf TCAD $N_\textrm{it,don}$}} & \multirow{2}{*}{{\bf TCAD $N_\textrm{it,acc}$}}\\[0.9mm]%
{\bf ID} & { } & [$\times10^{12}~\textrm{cm}^{-2}$] & [$\times10^{12}~\textrm{cm}^{-2}$] & [$\times10^{12}~\textrm{cm}^{-2}$] & [$\times10^{12}~\textrm{cm}^{-2}$]\\
\hline
{\bf N1(n)} & 
Inv./Acc. & $0.64\pm0.04$ & 0.68 & 1.11 & 1.08\\
\hline 
{\bf N2(n)} & 
Inv./Acc. & $0.77\pm0.10$ & 0.77 & 1.65 & 1.59\\
\hline
{\bf N3(n)} & 
Inv./Acc. & $1.06\pm0.15$ & 1.00 & 2.30 & 2.60\\
\hline
{\bf N4(p)} & 
Inv./Acc. & $2.00\pm0.14$ & 1.75 & 3.25 & 4.70\\
\hline
{\bf N5(n)} & 
Inv./Acc. & $2.1\pm0.3$ & 2.20 & 4.15 & 4.35\\
\hline
\multirow{2}{*}{{\bf G1(n)}} & 
Inv. & $2.47\pm0.15$ & 0.79 & 1.92 & 1.10\\
& Acc. & $2.19\pm0.14$ & 0.79 & 1.65 & 1.10\\
\hline
\multirow{2}{*}{{\bf G2(p)}} & 
Inv. & $1.66\pm0.19$ & 1.20 & 1.70 & 1.15\\
& Acc. & $2.17\pm0.20$ & 1.20 & 2.11 & 1.00\\
\hline
\multirow{2}{*}{{\bf G3(n)}} & 
Inv. & $3.2\pm0.4$ & 2.20 & 1.05 & 1.50\\
& Acc. & $2.4\pm0.4$ & 2.00 & 0.50 & 1.00\\
\hline
\end{tabular}
\end{table*}
%
\begin{table*}[!t]
\centering
\caption{The simulation input parameters of radiation-induced 
$N_\textrm{it}$. 
$E_\textrm{a,V,C}$ are the activation energy, 
valence band and conduction band energies, respectively, while $\sigma_\textrm{e,h}$ are the electron and hole trapping cross sections, respectively.} 
\label{tabNit}
\begin{tabular}{|c|c|c|c|}
    \hline
    {\bf $N_\textrm{it}$ type} & {\bf $E_\textrm{a}$} \textrm{[eV]} & {\bf $\sigma_\textrm{e,h}$} \textrm{[cm$^{2}$]} & 
{\bf Density} \textrm{[cm$^{-2}$]}\\
    \hline
    Deep donor ($N_\textrm{it,don}$) & $E_\textrm{V}+0.65$ & $1\times10^{-15}$ & see column 5 in Table~\ref{table_Simulated}\\
    Deep acceptor ($N_\textrm{it,acc}$) & $E_\textrm{C}-0.60$ & $1\times10^{-15}$ & see column 6 in Table~\ref{table_Simulated}\\
    \hline
\end{tabular}
\end{table*}

\begin{figure*}
     \centering
     \subfloat[]{\includegraphics[width=.49\textwidth]{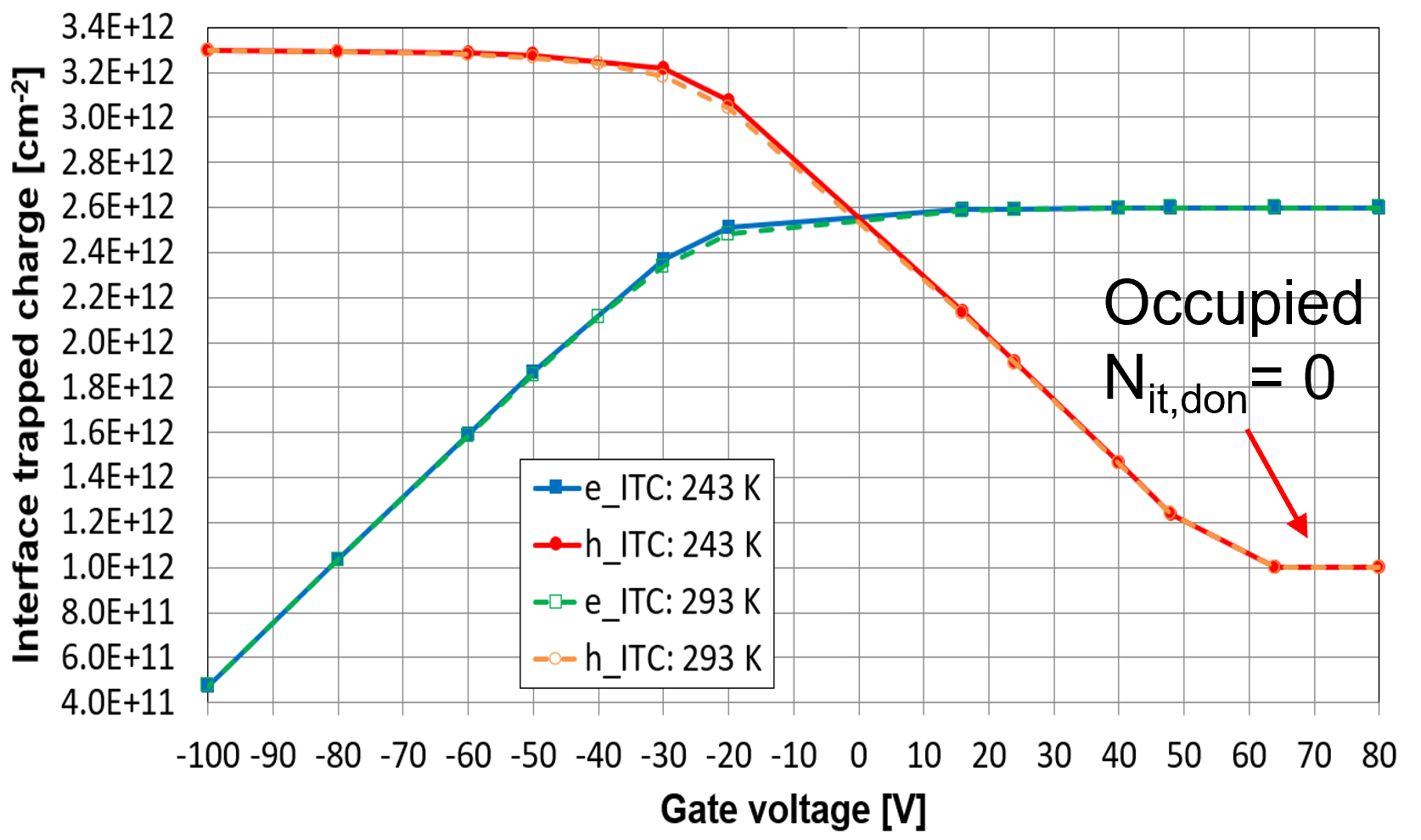}\label{120N_ehITC_235e15}}
     \subfloat[]{\includegraphics[width=.52\textwidth]{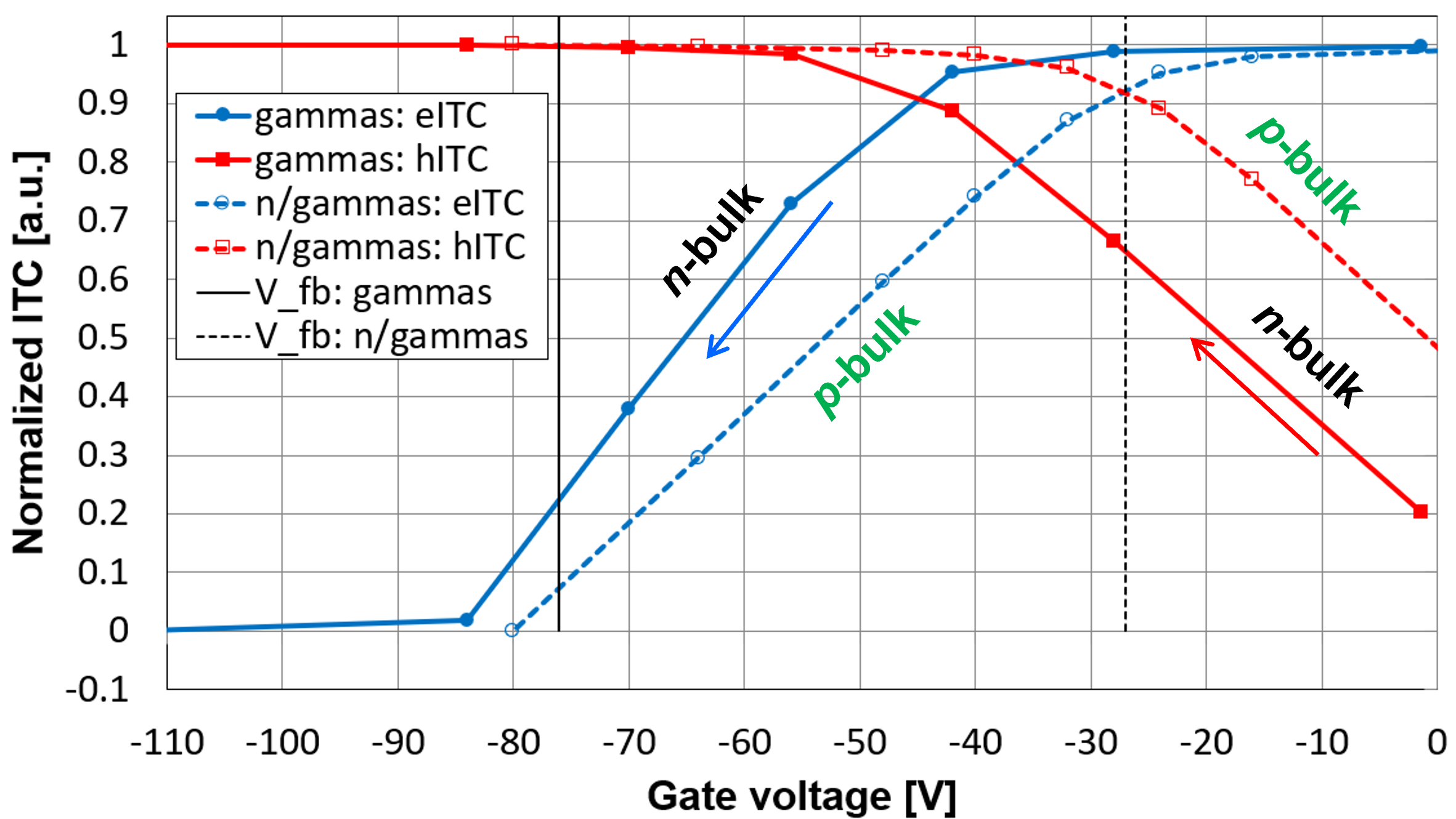}\label{ehITC_npBulk}}
    \caption{Simulated interface trapped charge densities 
    for (a) a 120-$\upmu$m-thick $p$-bulk (type inverted from $n$-bulk) MOS-capacitor at $T=$ +$20^{\circ}$ and -$30^{\circ}$ C, n/$\gamma$-irradiated to the fluence of $(2.35\pm0.19)\times10^{15}~\textrm{n}_\textrm{eq}\textrm{cm}^{-2}$ (from Figure~\ref{120N_4kHz_235e15}), and 
    (b) 300-$\upmu$m-thick $p$-bulk (type inverted from $n$-bulk) and 200-$\upmu$m-thick $n$-bulk MOS-capacitors at $T=$ +$20^{\circ}$ C, n/$\gamma$- and $\gamma$-irradiated to the fluence of $(6.1\pm0.5)\times10^{14}~\textrm{n}_\textrm{eq}\textrm{cm}^{-2}$ (MCNP-simulated $\textrm{TID}=7.1\pm0.6$ kGy) and dose of $7.0\pm0.4$ kGy, respectively (from Figures~\ref{300N_2kHz_61e14} and~\ref{200N_2kHz_7kGy}, respectively). The simulation input parameters of the $\gamma$-irradiated MOS correspond to the $CV$-curves starting from the accumulation-region in Figure~\ref{200N_2kHz_7kGy}.
   } 
\label{ITC}
\end{figure*}

\begin{linenomath}
For n/$\gamma$-irradiated MOS-capacitors, the accuracy of the simulation 
degrades at the highest positive $V_\textrm{gate}$-values in Figures~\ref{300N_1kHz_35e14},~\ref{300N_2kHz_61e14},~\ref{120N_4kHz_235e15} and~\ref{120N_3kHz_93e15}, where an increasing number of minority carriers are pulled to the Si/SiO$_2$-interface. Above $V_\textrm{th}$, the simulated $CV$-curve stabilizes to a constant value in the inversion region, which is not seen in the measurement. 
This is due to all $N_\textrm{it,don}$ becoming unoccupied, 
resulting in the net positive interface charge 
being composed solely of $N_\textrm{f}$ ($N_\textrm{f}=1.00\times10^{12}$ cm$^{-2}$ in row three of Table~\ref{table_Simulated}), as shown in Figure~\ref{120N_ehITC_235e15} where simulated interface-trapped charge density is displayed as a function of $V_\textrm{gate}$.
Thus, in the absence of $N_\textrm{M}$, the net oxide charge density can be described as
\begin{equation}\label{eq7}
Q_\textrm{ox}={e}N_\textrm{ox}=e[N_\textrm{f}+{a}N_\textrm{it,don}-{b}N_\textrm{it,acc}],
\end{equation}
where $a$ and $b$ are the fractions of fully occupied $N_\textrm{it,don}$ and $N_\textrm{it,acc}$, respectively. In Figure~\ref{120N_ehITC_235e15}, the values of $a$ and $b$ at -100 V are 1.00 and 0.18, respectively, while at +80 V the corresponding values are 0.0 and 1.0, respectively. This effectively makes $Q_\textrm{ox}$ in Eq.~\ref{eq7} a dynamic characteristic that experiences polarity change with $V_\textrm{gate}$.
\end{linenomath}

For $\gamma$-irradiated MOS in Figure~\ref{300P_4kHz_23kGy} the measured $C>C_\textrm{min}$ at deep inversion, showing signs of low-frequency measurement behavior where the inversion-layer charge is supplied/removed quickly enough to respond to changes in $V_\textrm{gate}$ (reducing depletion layer depth and increasing inversion $C$ towards $C_\textrm{ox}$) \cite{Nicollian1982}, which is not reproduced by the otherwise closely agreeing simulation at $f=4$ kHz. The other visible disagreement between measured and simulated $\gamma$-irradiated MOS-capacitors' $CV$-characteristics is seen in Figure~\ref{300N_3kHz_90kGy}, 
where the abrupt slope change at about 25 V into depletion-region after $V_\textrm{th}$ in measured $CV$-curve is not reproduced by the simulation. This could be evidence of an additional $N_\textrm{it}$ level introduced to the Si/SiO$_2$-interface at high doses, $N_\textrm{it,don}$ and $N_\textrm{it,acc}$ for $p$- and $n$-bulk MOSs, respectively, that becomes fully occupied at given $V_\textrm{gate}$-value and stops contributing dynamically to the shape of the depletion-region $CV$-slope. 
%
%

Using tuned values of $N_\textrm{f}$, $N_\textrm{it,don}$ and $N_\textrm{it,acc}$, it is possible to analytically find estimates for $a$ and $b$ in Eq.~\ref{eq7} that produce matching $N_\textrm{ox}$ with the measured values in Table~\ref{table_Simulated}. By plotting the TCAD-simulated evolution of peak-density-normalized fully occupied $N_\textrm{it,don}$ and $N_\textrm{it,acc}$, i.e. interface-trapped charge densities, with $V_\textrm{gate}$ in Figure~\ref{ehITC_npBulk}, it is possible to directly extract $a$ and $b$ at the point of $V_\textrm{fb}$. For $n$-bulk MOS G1(n) in Figure~\ref{ehITC_npBulk} $a=1.00$ and $b=0.22$, resulting in $N_\textrm{ox}=2.2\times10^{12}$ cm$^{-2}$ (with $N_\textrm{f}$, $N_\textrm{it,don}$ and $N_\textrm{it,acc}$ values 
for $CV$-sweep starting from accumulation in Table~\ref{table_Simulated}) that is within uncertainty of the measured $N_\textrm{ox}$. Corresponding values of $a$ and $b$ for MOS N2(n) in the same figure are 0.92 for both, resulting in $N_\textrm{ox}=0.825\times10^{12}$ cm$^{-2}$, which again is within uncertainty of the measured value. Thus, TCAD simulation and Eq.~\ref{eq7} provide a prediction of the fraction of fully occupied  $N_\textrm{it,don}$ and $N_\textrm{it,acc}$ at $V_\textrm{fb}$, where $N_\textrm{ox}$ is extracted.
%

\begin{figure*}
     \centering
      \subfloat[]{\includegraphics[width=.50\textwidth]{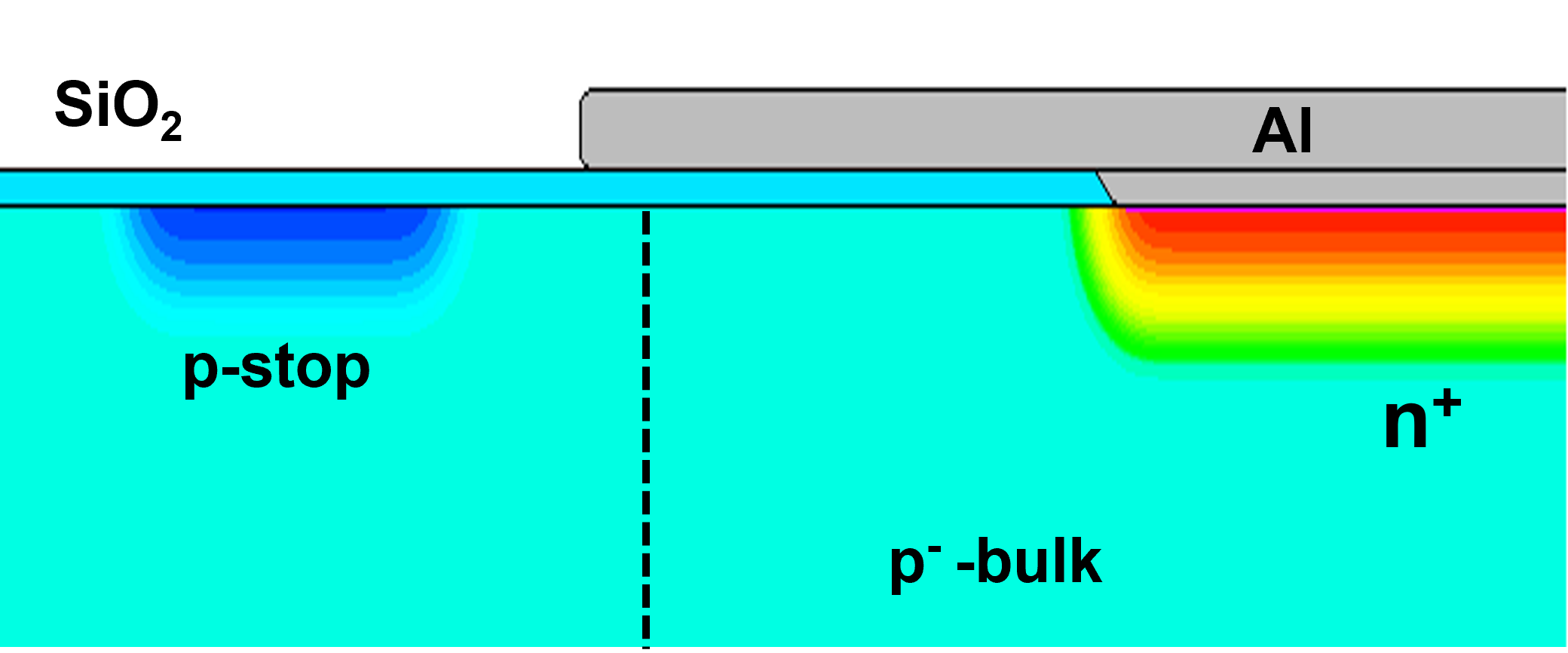}\label{intStrip}}\\
      \subfloat[]{\includegraphics[width=.50\textwidth]{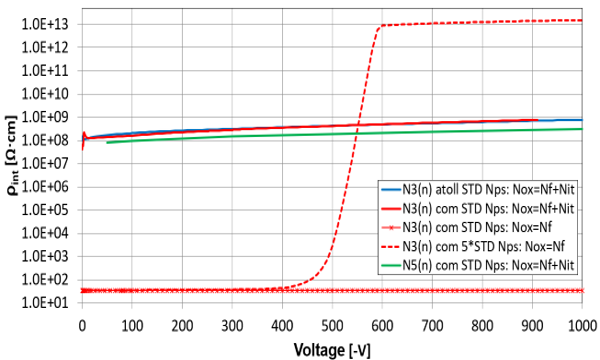}\label{Rint_Nit}}
      \subfloat[]{\includegraphics[width=.50\textwidth]{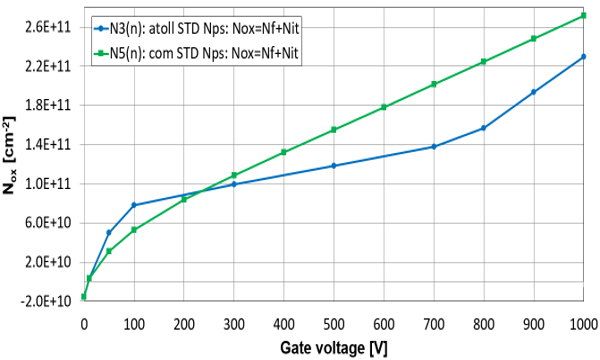}\label{Rint_NoxV}}
    \caption{Inter-strip simulations of a 300 $\upmu$m-thick $n$-on-$p$ sensor with 50 $\upmu$m strip-to-strip gap. (a) Close-up of the device structure from midgap to strip-implant with different layers and regions indicated. Dashed black line marks the location of the cut along the device surface in Figure~\ref{Rint_NoxV}. (b) Inter-strip resistivity evolution with reverse bias voltage at -$20^{\circ}$ C for individual (atoll) and common $p$-stop isolation implants 
($w_\textrm{ps}=6~\upmu\textrm{m}$, $d_\textrm{ps}=1.7~\upmu$m, standard value of $p$-stop peak doping: STD $N_\textrm{ps}=9\times10^{15}~\textrm{cm}^{-3}$), simulated by applying $N_\textrm{f}$ and $N_\textrm{it}$-parameters tuned from the measurements and simulations of MOS-capacitors N3(n) ($\Phi_\textrm{eff}=(2.35\pm0.19)\times10^{15}~\textrm{n}_\textrm{eq}\textrm{cm}^{-2}$) and N5(n) ($\Phi_\textrm{eff}=(9.3\pm1.1)\times10^{15}~\textrm{n}_\textrm{eq}\textrm{cm}^{-2}$). 
(c) $N_\textrm{ox}$ evolution with gate (strip) voltage at the Si/SiO$_2$-interface for the N3(n) 
with atoll $p$-stop and N5(n) in Figure~\ref{Rint_Nit}. 
}
\label{Rint_Nit_NoxV}
\end{figure*}

The 
$V_\textrm{gate}$-dependent dynamic properties of $Q_\textrm{ox}$ in Eq.~\ref{eq7} 
can potentially lead to improved isolation of segmented electrodes (e.g. pads, strips, pixels) in irradiated $n$-on-$p$ devices, that are in danger of getting shorted at high levels of positive polarity $N_\textrm{ox}$. This was investigated in Figure~\ref{Rint_Nit_NoxV} by using $N_\textrm{f}$, $N_\textrm{it,don}$ and $N_\textrm{it,acc}$ in Table~\ref{table_Simulated}, tuned to the measured $CV$-characteristics of the MOS-capacitors N3(n) (with $\Phi_\textrm{eff}=(2.35\pm0.19)\times10^{15}~\textrm{n}_\textrm{eq}\textrm{cm}^{-2}$) and N5(n) (with $\Phi_\textrm{eff}=(9.3\pm1.1)\times10^{15}~\textrm{n}_\textrm{eq}\textrm{cm}^{-2}$) as an input at the Si/SiO$_2$-interface for 
an inter-strip resistance ($R_\textrm{int}$)-simulation 
of a $n$-on-$p$ sensor structure in Figure~\ref{intStrip}. 
Displayed in Figure~\ref{Rint_Nit}, high levels of $R_\textrm{int}$ ($\rho_\textrm{int}$) are maintained throughout the investigated reverse bias $V$-range ($V_\textrm{bias}=0-1$ kV), for the `standard' value of $p$-stop peak doping (STD $N_\textrm{ps}$) and regardless of the $p$-stop configuration. Both the voltage dependence and absolute values of $\rho_\textrm{int}$ are in close agreement with previously reported experimental results of X-ray \cite{Moscatelli:2017sps} and reactor \cite{Dierlamm2017,Steinbruck2020} irradiated strip-sensors, while the gradual decay of measured $R_\textrm{int}$ with fluence is also reproduced. 
By excluding $N_\textrm{it,don}$ and $N_\textrm{it,acc}$ from the simulation, the strips 
become either shorted (STD $N_\textrm{ps}$, $N_\textrm{it}=0$ in Figure~\ref{Rint_Nit}) for all voltages or reach isolation only above 500 V of $V_\textrm{bias}$ 
for an extreme value of $p$-stop peak doping ($5\times\textrm{STD}$ $N_\textrm{ps}$, $N_\textrm{it}=0$). 

Figure~\ref{Rint_NoxV} shows the effect of fully occupied $N_\textrm{it}$ on $N_\textrm{ox}$ in the inter-strip region with the evolution of the positive potential at the strip electrode. For the MOS-capacitor N3(n), after polarity change from negative to positive between 0--10 V the values of $N_\textrm{ox}$ at the Si/SiO$_2$-interface only increase beyond pre-irradiated levels in Table~\ref{table_preIrrad} above 700 V, resulting in high $R_\textrm{int}$ throughout the $V$-range. Even though $N_\textrm{ox}$ keeps increasing above $2\times10^{11}~\textrm{cm}^{-2}$ (for N3(n) between 0--1 kV $a$ and $b$ in Eq.~\ref{eq7} change from 0.679 and 0.991 to 0.782 and 0.988, respectively), the corresponding $\rho_\textrm{int}$ in Figure~\ref{Rint_Nit} keeps growing, since electrons are getting increasingly swiped from the inter-strip region to the strip-electrodes at extreme voltages. The decay of $R_\textrm{int}$ for the higher fluence N5(n) from the levels observed for N3(n) in Figure~\ref{Rint_Nit}, is reflected by the higher values of $N_\textrm{ox}$ at $V>250$ V. Hence, the experimentally observed high $R_\textrm{int}$-values in irradiated $n$-on-$p$ strip-sensors can be explained by the beneficial effect to $R_\textrm{int}$ from the radiation-induced accumulation of $N_\textrm{it}$. Similar observations have been reported in previous simulation studies \cite{Moscatelli:2017sps,Gosewich2021} with varied approaches in the application of $N_\textrm{it}$. 

%% file: discussion_fT.tex
The $f$-dependence of the irradiated MOS-capacitor $CV$-characteristics is introduced by the capture and emission rates 
of $N_\textrm{it}$ \cite{Nicollian1982}, which is opposite for $N_\textrm{it,don}$ and $N_\textrm{it,acc}$. Thus, at increased $f$ 
a larger fraction of $N_\textrm{it,don}$ and $N_\textrm{it,acc}$ are fully occupied and unoccupied, respectively, leading to higher positive net oxide charge at Si/SiO$_2$-interface, 
while at decreased $f$ the trap occupation is reversed, resulting in lower positive $N_\textrm{ox}$ and therefore lower negative gate voltage 
produces the flat-band condition in both $p$- and $n$-bulk devices. 
Additionally, the Boltzmann statistics of deep-trap-level $e_\textrm{it}$ introduces significant temperature 
dependence ($e_\textrm{it}~{\propto}~e^{-E_\textrm{a}/kT}$, where $E_\textrm{a}$ is the activation energy of the trap) \cite{Nicollian1982}, 
when the capture and emission rates slow down with lower $T$ and measurements with constant $f$ result in substantially different $V_\textrm{fb}$ and $CV$-characteristics already within $\pm20^{\circ}$ C. These 
compromise the assessment of any fluence or dose-dependence of the Si/SiO$_2$-interface properties from MOS $CV$-characteristics, since the results are only valid for the given measurement $f$ and $T$. Frequency dependence changes also with the densities of $N_\textrm{it}$, thus no conclusions on the evolution of the oxide-interface properties between n/$\gamma$- and $\gamma$-irradiated MOS-capacitors or between different fluences and doses are included in this study. 

%% file: Summary.tex
Combined experimental and simulation $CV$-characterization study of MOS-capacitor samples diced from 6-inch wafers --with about 700-nm-thick gate oxides-- suggests that $N_\textrm{ox}$ at the Si/SiO$_2$-interface changes from being dominated by $N_\textrm{f}$ before irradiation to include both deep donor- ($E_\textrm{V}+0.65$ eV) and deep acceptor-type ($E_\textrm{C}-0.60$ eV) $N_\textrm{it}$, with comparable densities to $N_\textrm{f}$, after being exposed to either a $\gamma$-source or to a reactor radiation environment. This was found by the process of reproducing the measured $CV$-characteristics of five n/$\gamma$-irradiated and three $\gamma$-irradiated MOS-capacitors by a TCAD-simulation, where close agreement between measurement and simulation were only reached when in addition to $N_\textrm{f}$, both $N_\textrm{it,don}$ and $N_\textrm{it,acc}$ were included in the tuning.

Due to the $V_\textrm{gate}$-dependence of the fraction of the fully occupied $N_\textrm{it,don}$ and $N_\textrm{it,acc}$, the dynamic nature of $N_\textrm{ox}$ can 
explain the experimentally observed high levels of inter-electrode resistance in irradiated $n$-on-$p$ devices, like strip-sensors. An $R_\textrm{int}$-simulation of a $n$-on-$p$ strip-sensor structure using $N_\textrm{f}$, $N_\textrm{it,don}$ and $N_\textrm{it,acc}$ (tuned to reproduce the $CV$-characteristics of an irradiated MOS-capacitor) as an input, 
showed close agreement for both the voltage dependence and absolute values of $R_\textrm{int}$ 
with previously reported experimental results of X-ray and reactor irradiated strip-sensors. High values of simulated $R_\textrm{int}$ were due to the low levels of $N_\textrm{ox}$ in the inter-strip region between the strip-anodes.

Thus, the simulation study provides two separate observations of the macroscopic effects of the radiation-induced accumulation of $N_\textrm{it}$ at the Si/SiO$_2$-interface that can explain the experimental results. The $N_\textrm{it}$-accumulation-generated beneficial impact on 
inter-electrode isolation in irradiated $n$-on-$p$ sensors suggests that very high levels of $p$-stop peak doping ($\gtrsim5\times10^{16}~\textrm{cm}^{-3}$) could be avoided, mitigating the probability of discharges or avalanche effects due to excessive electric fields at the $p$-stops in the extreme radiation environment of the foreseen high-luminosity LHC. 